# Local Indicator of Colocation Quotient with a Statistical Significance Test: Examining Spatial Association of Crime and Facilities[1]

Fahui Wang[1*], Yujie Hu[1], Shuai Wang[2] and Xiaojuan Li[2]

[1]Department of Geography & Anthropology, Louisiana State University, Baton Rouge, LA 70803, USA
[2]College of Resources, Environment & Tourism, Capital Normal University, Beijing 100048, China

**Abstract**
Most existing point-based colocation methods are global measures (e.g., join count statistic, cross K function, and global colocation quotient). Most recently, a local indicator such as the local colocation quotient is proposed to capture the variability of colocation across areas. Our research advances this line of work by developing a simulation-based statistic test for the local indicator of colocation quotient (LCLQ). The study applies the indicator to examine the association of land use facilities with crime patterns. Moreover, we use the street network distance in addition to the traditional Euclidean distance in defining neighbors since human activities (including facilities and crimes) usually occur along a street network. The method is applied to analyze the colocation of three types of crimes and three categories of facilities in a city in Jiangsu Province, China. The findings demonstrate the value of the proposed method in colocation analysis of crime and facilities, and in general colocation analysis of point data.

## Introduction

The spatial relationship between objects has long been of great interest to geographers. One type of research on this theme is *spatial autocorrelation* analysis that measures the association between objects of the same type (Cliff and Ord 1973). Existing measures include global (Moran 1950; Geary 1954) and local statistics (Getis and Ord 1992; Anselin 1995) with wide applications such as crime analysis (e.g., Messner et al. 1999; Getis 2010) and epidemiology (e.g., Moore and Carpenter 1999). Another line of research focuses on the so-called *spatial correlation* or *colocation*, which measures the spatial relationship between objects of different types, another fundamental relationship in geographic data (Leslie and Kronenfeld 2011).

---

[1] This is a preprint of: Wang, F., Hu, Y., Wang, S., & Li, X. (2017). Local indicator of colocation quotient with a statistical significance test: examining spatial association of crime and facilities. *The Professional Geographer*, *69*(1), 22-31. https://doi.org/10.1080/00330124.2016.1157498

Commonly used techniques of colocation analysis can be grouped into two categories: area-based and point-based approaches. For example, Anselin (2003) designed a bivariate local indicator of spatial association (LISA) to investigate the spatial correlation pattern (e.g., high-high, low-low, high-low, and low-high) between two area-based variables (e.g., number of alcohol outlets vs. number of crime incidents in areas), and provided an assessment of statistical significance as well. Join (or Joins or Joint) count statistic is another area-based measure of spatial correlation, but is only limited to data of binary values (Dacey 1965). The method basically examines the association between two categories of polygons by first counting the number of links whose two end-polygons are of different categories and then comparing the counts with expected counts by chance (Cliff and Ord 1981). The statistic is direction-blind, i.e., it does not differentiate the counts of A→B and B→A joins (Leslie and Kronenfeld 2011).

For point-based colocation analysis, the cross K (or bivariate K) function—an extension of Ripley's K function (Ripley 1976, 1977) for two populations—is frequently used to investigate the relationship between two spatial processes under the assumption of independence (e.g., Groff, Weisburd, and Morris 2009; Groff, Weisburd, and Yang 2010). It measures the ratio of observed overall density of type B points within a specified distance (usually in Euclidean distance) of a type A point over what we would expect by chance (Cressie 1991; Cuthbert and Anderson 2002). Okabe and Yamada (2001) then improved the measure by using network distance since urban activities (e.g., crimes) usually occur along street network (Krause 2012). Leslie and Kronenfeld (2011) found that the colocation association detected by the cross K function would be highly biased by the distribution pattern of the underlying population (e.g., clustering). They went on to propose a *colocation quotient (CLQ)* measure based on the concept of location quotient in economics (Isserman 1977) as well as a statistical test procedure. The CLQ is a global measure of colocation pattern, and thus cannot detect how the strength of association changes from place to place. More recently, Cromley, Hanink, and Bentley (2014) developed a local version of CLQ, here termed *local indicator of colocation quotients (LCLQ)*, by integrating the concept of geographically weighted parameters in spatial regression models with the global CLQ. However, no statistical test is proposed for it, and thus one cannot identify where a colocation is statistically significant. This article advances this line of work by developing a simulation-based statistical test for the LCLQ.

To illustrate our approach, it is applied to analyze the colocation of three types of crimes and three categories of facilities in a city in Jiangsu Province, China. Understanding the colocation patterns between crimes and facilities may help answer why crime happens where it does, and guide place-based policing strategies. For example, some facilities are susceptive of spurring higher crime intensity at and around them such as off-premise alcohol outlets (Scribner et al. 1999; Grubesic and Pridemore 2011; Pridemore and Grubesic 2012), on-premise alcohol outlets (Liang and Chikritzhs 2011; Day et al. 2012; Groff 2013), or underground transit stations (Ceccato and Uittenbogaard 2014). To measure their colocation associations, researchers commonly used methods such as a simple count approach (e.g., the number of facilities within a threshold distance of crimes), its variants by considering the distance decay effect or the activity intensity at facilities, the cross K function, and the bivariate LISA. However, as previously pointed out, these methods measured crime by a ratio or interval variable and were biased because of the joint distribution pattern (e.g., clustering) of population and crime. Our proposed LCLQ with a statistical significance test mitigates the concern and provides a more reliable approach to detecting colocation behavior and its locations. The contributions of this research include proposing a significance test for the LCLQ that is important but absent in the literature and using the network distance in defining CLQ besides Euclidean distance.

**Study area and data sources**

The study area is a city in Jiangsu Province, China (the name of the city is not identified due to our user agreement with the data provider). According to the census data in 2010, the city has 4.59 million "permanent residents" (with registered residence status) and 1.33 million "temporary residents" (i.e., migrant workers and their family members). As shown in Figure 1A, it consists of four districts (known as *qu* in Chinese, labeled as District A, B, C, and D here) and 25 subdistricts (known as *jie-dao* in Chinese, the smallest geographic unit for census data available to the public in China). District A, with a population of 456,000, in the north is established as a high-tech manufacturing area with about a thousand domestic and overseas companies. District B, with a population of 367,200, in the west is an area with mainly commercial, tourist, and real estate development companies with a dense road network. For example, the largest unofficial central business district (CBD) is located on the east edge of this district, adjacent to District C. Similarly, District C has many commercial and tourist activities with a population of 450,100. Most of the city's traditional manufacturing and energy production industries are concentrated in District D, with a population of 88,100. Overall, District B and C (especially the areas between them) are highly developed.

Crime data, provided by the city's Bureau of Public Safety, include three types of crime with the highest numbers of incidents, namely residential burglary, robbery, and motorcycle theft, for about two years (October 2009 – August 2011). Each crime incident has its occurrence time and location (Figure 1B). Other supporting data include the road network, locations of major land use facilities such as schools, retail shops and entertainment establishments (Figure 1C). There were 7,777 motorcycle theft, 5,677 residential burglary, and 429 robbery incidents recorded during the study period, and most of them occurred in the south part, especially District B and C. Crime counts are considerably lower as recorded by police in Chinese cities than western cities. Major land use facilities selected for this study included 143 schools, 215 large retail shops, and 450 entertainment establishments. Entertainment establishments consisted of internet cafés, amusement arcades, leisure centers, nightclubs, and karaoke clubs.

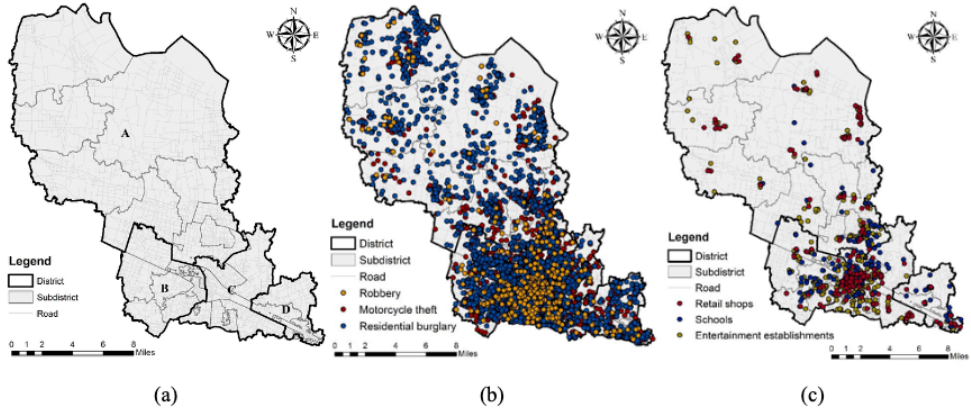

Figure 1. (a) a city in Jiangsu Province, China; (b) three types of crimes in the city; (c) three types of facilities in the city

**Colocation analysis by cross K function**

As the baseline analysis for comparison, the traditional cross K (also known as bivariate K) function is used to test for independence between crimes and land use facilities in the study area. As an extension of Ripley's K function for two populations, this technique measures the ratio of observed overall density of type B points in proximity to type A points with what would be expected by chance (Cressie 1991). Specifically, it compares the overall number of type B points within a specified threshold distance (either Euclidean or network distance) around each type A point with the number in a random model (Leslie and Kronenfeld 2011). A significance test is fulfilled by a series of Monte Carlo simulation. Similar to the output of Ripley's K function, results for cross K function analysis are presented as a graph illustrating the change of cross K value ($y$-axis) as distance ($x$-axis) increases. Based on the graph, one can detect at what distance range the two types are spatially correlated (i.e., colocated) or independent from each other. For example, statistically significant colocation exists at a certain distance range if the cross K value at that distance falls above an upper envelop returned by the significance test. Similarly, significant dispersion of two point sets is suggested if the cross K value line lies below the lower envelop line as calculated by the significance test. If a cross K value falls within the upper and lower envelops, the spatial distribution of type A points has no impact on that of type B points (Groff, Weisburd, and Morris 2009).

For implementation, we used the Ripley's cross K function provided in GeodaNET (https://geodacenter.asu.edu/downloads/software/gnet) to analyze the colocation associations of crimes and major facilities. Figures 2A shows the result of analysis between motorcycle theft incidents and entertainment establishments based on Euclidean distance. The pattern for other two crime types are similar and are thus not shown. In general, all three types of crimes significantly colocated with entertainment establishments at a straight-line distance no more than 33 km. In other words, the spatial distribution patterns of all three types of crimes were significantly affected by the distribution pattern of entertainment establishments. Significant colocation associations are also detected between crimes (all three categories) and two other facilities (namely, schools and major retails). Results are not shown here.

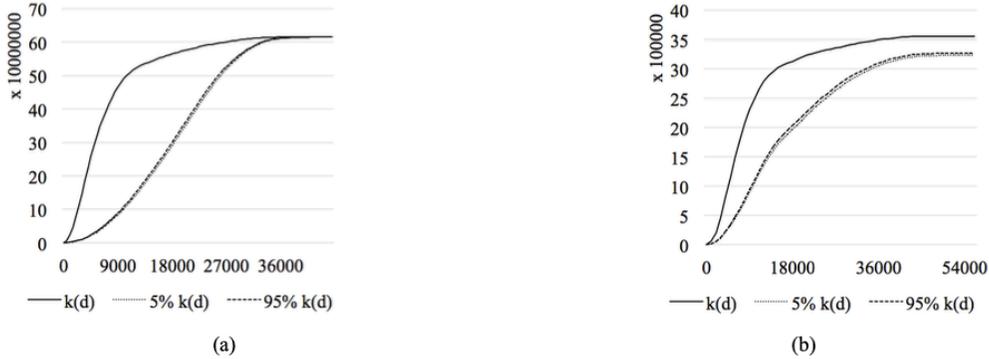

Figure 2. Cross K function analysis of motorcycle theft and entertainment establishments [$x$-axis shows distance in meters and $y$-axis represents cross K value]: (a) Euclidean-distance-based; (b) Network-distance-based

Different from the join count statistic, the cross K function is sensitive to the direction of colocation association. However, it may not make much sense here to analyze the colocation behavior from land use facilities to crimes, i.e., the degree to which one type of land use facility is attracted to a certain category of crime. Thus, the other direction of colocation is not explored here and hereafter.

**Analysis by global colocation quotient (CLQ)**

It is rather unsurprising that the above cross K function analysis suggests colocation between crimes and land use facilities. As Leslie and Kronenfeld (2011) pointed out, the colocation estimates of two point populations may be significantly biased by the distribution of the underlying population if they collectively exhibit a significant spatial clustering pattern. In our case, the data of the joint population (i.e., crime and land use facility combined) in the study area display a statistically significant clustering pattern. According to the nearest neighbor index, its value is 0.41 with a Z score of -137.89.

To mitigate the bias, Leslie and Kronenfeld (2011) proposed the colocation quotient (CLQ), which is based upon the concept of location quotient commonly used in economic geography studies (Stimson, Stough, and Roberts 2006). The CLQ examines the overall association between observed and expected numbers of type B points in proximity to type A points. In this regard, this indicator is termed the *global CLQ*. Different from the cross K function, the global CLQ is designed on the basis of nearest neighbors rather than actual metrical distance in order to account for the aforementioned effect of the joint population distribution.

The global CLQ is formulated as

$$CLQ_{A \to B} = \frac{N_{A \to B}/N_A}{N_B/(N-1)} \tag{1}$$

where $N$ represents the total number of points, $N_A$ denotes the population size of type A points, $N_B$ depicts the number of type B points, $N_{A \to B}$ denotes the number of type A points that have type B points as their nearest neighbors. The numerator calculates the *observed* proportion of type B points that are the nearest neighbors of type A points, while the denominator estimates the *expected* proportion by chance. Given that a point itself is not an eligible candidate of its nearest neighbor, $N$-1 rather than $N$ is used in the measure of expected proportion. As mentioned previously, the cross K function takes the direction of colocation relationship into account, and likewise, the global CLQ can distinguish the spatial interaction from A to B from the opposite. In other words, $CLQ_{A \to B}$, which stands for the extent to which type A points are attracted to type B points, differs from $CLQ_{B \to A}$ that indicates to what degree type B points are drawn by type A points.

In practice, a point A may have multiple nearest neighbors within a bandwidth. Under this circumstance, the global CLQ treats each nearest neighbor of point A equally in the calculation of $N_{A \to B}$. It is formulated as

$$N_{A \to B} = \sum_{i=1}^{N_A} \sum_{j=1}^{nn_i} \frac{f_{ij}}{nn_i} \tag{2}$$

where $i$ denotes each type A point, $nn_i$ represents the number of nearest neighbors of point $i$, $j$ indicates each of point $i$'s nearest neighbors $nn_i$, and $f_{ij}$ is a binary variable indicating whether or not point $j$ is of type B under investigation (1 indicates yes and 0 otherwise).

The global CLQ is proved to have an expected value of one when all points are randomly re-labeled given the frequency distribution of each point set (Leslie and Kronenfeld 2011). Consequently, a global $CLQ_{A \to B}$ larger than one indicates that type

A points and B points are spatially colocated. To be more specific, type A points tend to be more attracted to B points in space, and a larger CLQ value indicates a stronger colocation association. On the contrary, a global $CLQ_{A \to B}$ less than one suggests that type A points tend to be isolated from B points. To determine whether or not such spatial colocation or isolation behavior is significantly nonrandom, a Monte Carlo simulation-based statistical test is followed to compare with the null hypothesis. In the cross K function, the null hypothesis is no spatial association between the pair of point sets under investigation. However, the global CLQ examines if this assumption still holds given the distribution pattern of the joint population (e.g., clustered or dispersed in general). This again indicates that the global CLQ can detect the colocation association free from the distribution pattern of the underlying population. The Monte Carlo simulation process reassigns the category of each point in a random manner, but subject to the frequency distribution of each point category. A sampling distribution in terms of global CLQ is obtained by repeating this process for many times. Then the distribution of observed global CLQ is compared with the sampling distribution to derive a test statistic and significance level.

Leslie and Kronenfeld (2011) developed a tool to implement the global CLQ method (http://seg.gmu.edu/clq). Following their approach, we set the bandwidth as the 1[st] order neighbor (i.e., the nearest neighbor). The significance test was conducted by running the Monte Carlo simulation for 1,000 times. Table 1 lists the global CLQs of crimes vs. land use facilities, or more specifically, the overall degree of crimes attracted to land use facilities. Results show that the numbers of any land use facilities as the nearest neighbors of any type of crimes were less than expected by chance, and crimes—regardless of types—were significantly (at 0.001 level) isolated from any of the three facility types in the study area. This contradicts the finding by the cross K function analysis. That is to say, the finding from the cross K function is likely to be attributable to the significant clustering pattern of the joint population, and can be biased or even erroneous.

Table 1. Global CLQ of crimes vs. land use facilities

| Crime type | Schools | | Retail Shops | | Entertainment establishments | |
|---|---|---|---|---|---|---|
| | bandwidth=1 | bandwidth=10 | bandwidth=1 | bandwidth=10 | bandwidth=1 | bandwidth=10 |
| Robbery | 0.601 | 0.881 | 0.590 | 0.815 | 0.686 | 0.923 |
| Residential burglary | 0.666 | 0.732 | 0.483 | 0.679 | 0.475 | 0.679 |
| Motorcycle theft | 0.567 | 0.660 | 0.442 | 0.655 | 0.496 | 0.681 |

Note: "bandwidth=1" uses 1 nearest neighbor, "bandwidth=10" uses 10 nearest neighbors; all CLQ values are significant at the 0.001 level.

**Analysis by local indicator of colocation quotient (LCLQ)**

Both global measures of colocation relationship in the cross K function and global CLQ implicitly assume that the (in)dependent association stays stationary over space. However, such a spatial correlation is likely to change from place to place, just like the parameters in a geographically-weighted regression model (Fotheringham, Brunsdon, and Charlton 2003) or some local spatial autocorrelation indices such as the $G_i$ statistic (Getis and Ord 1992) and the LISA (Anselin 1995). Therefore, a local version of CLQ is proposed by Cromley, Hanink, and Bentley (2014) to reveal the spatial variability of correlation between two point sets. This *local indicator of colocation quotient*, abbreviated as *LCLQ*, may help us better understand the spatial process and the underlying driving factors. In addition, the LCLQ can provide mappable and more readable results.

The LCLQ is formulated as

$$LCLQ_{A_i \to B} = \frac{N_{A_i \to B}}{N_B/(N-1)} \quad (3)$$

$$N_{A_i \to B} = \sum_{j=1(j \neq i)}^{N} \left( \frac{w_{ij} f_{ij}}{\sum_{j=1(j \neq i)}^{N} w_{ij}} \right) \quad (4)$$

$$w_{ij} = \exp(-0.5 * \frac{d_{ij}^2}{d_{ib}^2}) \quad (5)$$

where the LCLQ for point $A_i$ relative to type B points has a similar expression as the global CLQ in Equation 1, $N_{Ai \to B}$ denotes the weighted average number of type B points that are the nearest neighbors of point $A_i$, $f_{ij}$ still represents a binary variable indicating if point $j$ is a marked B point (1 for yes and 0 otherwise), $w_{ij}$ denotes the weight of point $j$, indicating the importance of point $j$ to the $i$th A point, $d_{ij}$ is the distance between the $i$th A point and point $j$, $d_{ib}$ denotes the bandwidth distance around the $i$th A point. Equation 5 defines the Gaussian kernel density function that is used to assign geographic weights to each neighbor of point $A_i$, and basically the farther a neighbor locates beyond point $A_i$, the less important it is to point $A_i$. One may also consider other density functions to define a point's weight, such as the so-called box kernel density function which treats each point within the prescribed bandwidth distance from a marked A point identically as one, regardless of their distances (Cromley, Hanink, and Bentley 2014).

Similar to the implementation of global CLQ, the LCLQ also adopts distance ranks (e.g., the $1^{st}$ nearest neighbor or the $2^{nd}$ nearest neighbor) instead of actual distance metric as the bandwidth in calculation. This bandwidth setting is also known as *adaptive bandwidth*. Compared to a fixed bandwidth based on metrical distance, the adaptive bandwidth ensures that exactly the same number of points is involved in the estimation of LCLQ at each marked A point, and thus returns more robust and reliable results. In practice, the order of neighbors (e.g., the $1^{st}$ nearest neighbor and the $2^{nd}$ nearest neighbor) may be estimated by different distance measures such as the Euclidean and network distances. The Euclidean distance is computationally efficient, but the network distance measure is more accurate especially in an area with wide variability of road network. We suspect that the difference in results from the two distance measures may be minor in developed urban areas, but significant in suburban and rural areas. Such a potential variation, however, is not considered in the work on the global CLQ by Leslie and Kronenfeld (2011) or the LCLQ by Cromley, Hanink, and Bentley (2014).

As illustrated in Cromley, Hanink, and Bentley (2014), the LCLQ could offer more insights on the spatial colocation between two point sets than the global version. However, its value is limited if we do not know whether the detected colocation is statistically significantly different from random. This article further extends their work by incorporating a rigorous statistical test. Similar to Leslie and Kronenfeld (2011), the Monte Carlo simulation technique is adopted to generate sample distribution. In spatial analysis, this technique is often used as a tool for statistical test instead of a simple statistic such as the *t*-test, because it does not make any assumption about the expected distribution (Shi 2009). At a given location, one simulation trial randomly reassigns the group (category) of any point other than the given point itself (see Kronenfeld and Leslie (2015) for more details on the so-called *restricted random labeling* approach), and the frequency distribution of each category is kept fixed in this relabeling step. For example, if there are totally four point sets—A, B, C, and D—under study, the simulation process at point $i$ re-labels any other points in a random fashion until all points are re-categorized; and the number of elements in each point set remains unchanged after the simulation. Normally, this simulation trial is repeated for a predetermined number of times, and a sample distribution of LCLQs can be obtained at each point of interest (say, point $i$) by recalculating a LCLQ for each trial. Then the significance of calculated LCLQ at point $i$ is determined by comparing it with the sample distribution generated from simulation. Specifically, we may calculate the two-tailed significance by doubling the minor one between the proportions of simulation trials that have a recalculated LCLQ no less than the observed LCLQ and that have a simulated LCLQ no larger than the observed LCLQ (Leslie and Kronenfeld 2011).

We developed a program coded in C# to implement calibration of the aforementioned LCLQ and corresponding statistical test. For illustration purposes, we only show the results between three types of crimes and one type of facility (entertainment establishments). Different from results of global CLQ listed in Table 1, the bandwidth here is set to be 10 nearest neighbors around each examined point for a smoother and more readable pattern. We will discuss more on the bandwidth selection in the next section. Before the local colocation analysis, we first calculate the global CLQs of crimes vs. land use facility based on a bandwidth of 10 nearest neighbors, which may serve as a benchmark to better understand and interpret the LCLQs. Also listed in Table 1, all global CLQs regardless of the combination of crime and facility experienced a significant increase relative to results when the bandwidth was 1. This may indicate our suspicion of the smoothing role of a greater bandwidth value. After increasing the bandwidth, most global CLQs are still well below 1 (e.g., residential burglary and motorcycle theft); however, the CLQ value for robbery vs. entertainment establishments is close to 1, indicating a spatial association of less dispersion between robbery incidents and entertainment establishments.

Figures 3A-3C illustrate the Euclidean-distance-based local colocation results between three types of crimes and entertainment establishments. Based on Figure 3A, among the robbery incidents significantly nonrandom at 0.05 level, almost all have a LCLQ greater than one, and they are concentrated in the bordering area between Districts B and C. Robbery incidents are significantly colocated with entertainment facilities in these areas near the CBD where social networking activities congregate in major entertainment establishments. Some colocations between robbery and entertainment establishments are also observed in the southeast corner of District A where several parks and a major university campus are located. In terms of residential burglary and motorcycle theft, the global colocation analysis finds a significant dispersion relationship with any land use facilities (at 0.001 level). However, the LCLQs detect some areas where crimes are significantly colocated with entertainment establishments. As shown in Figure 3B, residential burglary incidents colocate with entertainment facilities significantly (1) mainly around but not in the core of the CBD area (probably due to fewer residential communities relative to commercial establishments in the core of the CBD area), (2) in the southeast corner of district A where parks and a university campus were located, and (3) in several small communities in the peripheral and middle of District A. Based on Figure 3C, significant colocations between motorcycle theft incidents and entertainment facilities are mostly observed in the CBD area and the southeast corner of district A, and are scattered in small pockets in the north of District A and the southeast corner of District D. The above findings indicate that the global CLQ results—significant dispersion pattern between crimes and facilities, regardless of

types—do not hold from place to place, and thus demonstrate the usefulness of LCLQ, especially for areas with heterogeneous land use patterns.

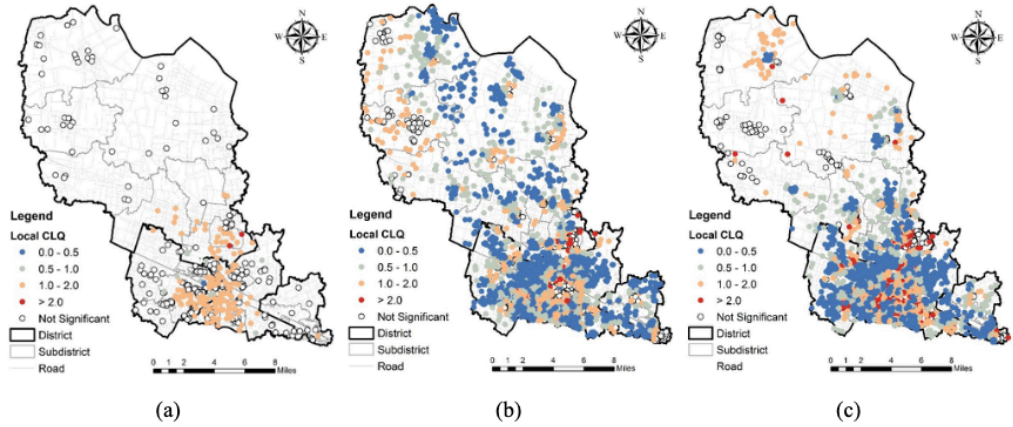

Figure 3. Euclidean-distance-based LCLQ between crimes and entertainment establishments [adaptive bandwidth = 10 nearest neighbors, 0.05 as threshold significance level]: (a) robbery; (b) residential burglary; and (c) motorcycle theft

**Discussion**

The choices of some parameters or metrics in implementing the LCLQ calibration may affect the results. This section examines the impacts of bandwidth and the distance metric.

Bandwidth is a critical parameter in kernel density-based methods (Anderson 2009). So is our case in both the global CLQ and LCLQ. Smaller and spiky clusters are expected when a smaller bandwidth is used, and change to smoother and larger clusters when a larger bandwidth is used (Hu, Miller, and Li 2014). Here we use an exploratory experiment with three different bandwidth values (i.e., $1^{st}$, $10^{th}$, and $25^{th}$ order neighbors) to investigate the impact of an increasing bandwidth on the CLQ analysis. One may consider adopting some bandwidth selection methods such as the data-based selection (Sheather and Jones 1991), cross-validation (Brunsdon 1995), and distance-based approaches (Fotheringham, Brunsdon, and Charlton 2000) for an appropriate bandwidth estimate.

Figure 4 illustrates the results of global CLQ between motorcycle thefts and each of three facility types. As we increase the search bandwidth from the $1^{st}$ order neighbors to the $10^{th}$ and further to the $25^{th}$, the global CLQs between motorcycle thefts and any facilities (regardless of facility type) become greater, but still less than one. In other words, motorcycle thefts are distributed in isolation from three examined land use facilities in the study area, but a greater bandwidth detects a dispersed association of less extent. In short, a small bandwidth returns a spiky or less smoothed result, and a large bandwidth (e.g., 25 nearest neighbors) yields an over-smoothed outcome. A bandwidth of 10 nearest neighbors may be a reasonable choice.

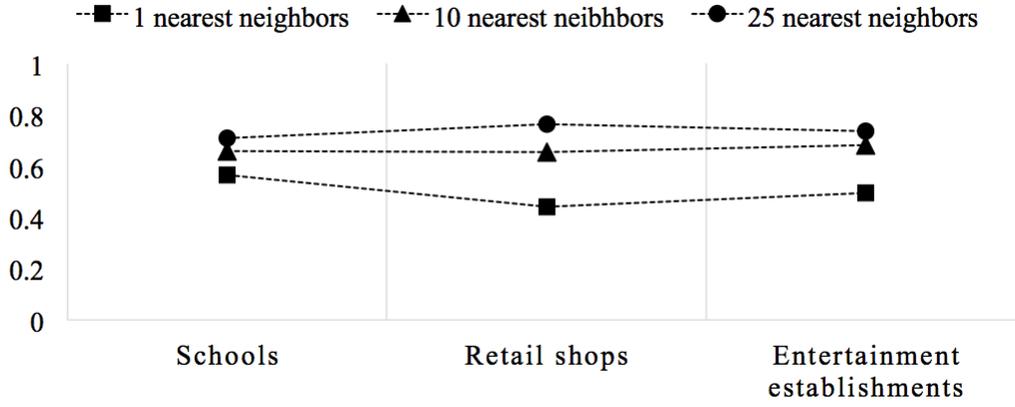

Figure 4. Global CLQs between motorcycle thefts and facilities with three bandwidth values [All CLQs are significant at 0.001 level]

For the LCLQ analysis, only the results from motorcycle thefts (vs. any facilities) are presented here as motorcycle thefts have the largest sample size. Figures 5A-B show the LCLQ results for bandwidths of 1 and 25 nearest neighbors, respectively. Note that the result for bandwidth of 10 is already shown in Figure 3C. In general, the results get smoother as the bandwidth increases. A small bandwidth of 1 nearest neighbor provides a less smoothed result of a few spiky 'spots'—with a significant colocation association with entertainment establishments—surrounded by a large number of areas that are significantly dispersed from entertainment establishments. A large bandwidth of 25 nearest neighbors, however, generates a result of several large clusters of motorcycle theft incidents that are significantly attracted to entertainment establishments but far fewer incidents that are isolated from the facilities. In the meantime, the number of incidents with insignificant LCLQ scores declines as the bandwidth increases. Again a bandwidth of 10 provides a reasonable balance between local details and general regional trend and thus is recommended.

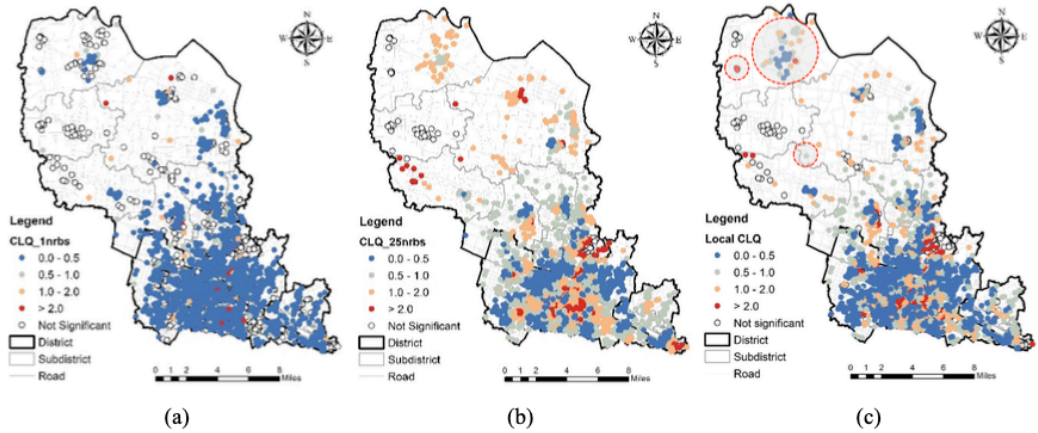

Figure 5. Euclidean-distance-based LCLQs between motorcycle thefts and entertainment establishments with a bandwidth of: (a) 1 and (b) 25 nearest neighbors [See Figure 2(c) for a bandwidth of 10]; (c) Network-distance-based LCLQ between motorcycle thefts and entertainment establishments with an adaptive bandwidth of 10 nearest neighbor

As pointed out previously, road network distance may be preferred to Euclidean distance in measuring spatial impedance as urban activities such as crimes and facilities are mostly confined (and possibly even geocoded) to the existing street network. Figure 2B shows the cross K function result between motorcycle thefts and entertainment establishments based on network distances. The observed cross K value is significantly higher than the upper envelop curve across all distance ranges (from 0 to 54 km) in the network distance-based measure. The result by Euclidean distances, as reported in Figure 2A, shows that the detected colocation association is not significant beyond 33 km. Therefore, a stronger colocation association is detected by the cross K function analysis based on network distance than that based on Euclidean distance.

We developed a program in C# to implement the network distance-based global CLQ. Again, using the global CLQ on motorcycle thefts vs. entertainment establishments with a bandwidth at 10 nearest neighbors as an example, the global CLQs measured by Euclidean distance and network distance are 0.681 and 0.686 (both significant at 0.001 level), respectively. In other words, adopting a more reasonable distance metric does not change the global CLQ much in our case study.

How does distance metric affect the local CLQ? On the same example (i.e., LCLQ on motorcycle thefts vs. entertainment establishments with a bandwidth at 10), a simple correlation analysis reveals that the network distance-based LCLQs are highly correlated with the Euclidean distance-based results with a correlation coefficient of 0.86 (significant at 0.001 level). That is to say, the LCLQ results based on two distance metrics are largely consistent with each other. However, there are significant discrepancies between the two. Figure 5C presents the result of network distance-based LCLQ. Comparing Figure 5C to Figure 3C for the Euclidean distance-based LCLQ, one can spot significant variations in the northwest corner of the study area as marked by the dashed circle, but not in the south (e.g., southeast corner of District A and most part of District B, C and D). For example, the larger circle in the northwest corner of the study area highlights a cluster of nine motorcycle theft crime incidents that are reported to significantly colocate with entertainment establishments with an average network distance-based LCLQ of 2.15, while insignificant colocation with an average Euclidean distance-based LCLQ of 3.06. Such a difference may be explained by the road network pattern in that area. Specifically, there are only a few roads connecting this area with others and the road density around that area is very low; therefore, the distance between an incident and its 10 nearest neighbors would become much longer as measured by network distance than the Euclidean distance, and thus would return results with lower LCLQ values but higher confidence. Similarly, we also find lower LCLQ values by the network distance measure than the Euclidean distance approach in other two areas (in the two marked smaller circles in Figure 5C) where road network is also sparse. This may suggest that the network distance-based LCLQ approach, in general, reports lower but more significant LCLQ values than the Euclidean distance approach, especially in the rural area where road network is sparse and simple. This finding is consistent with that from the network distance-based cross K function as shown in Figure 2B.

**Conclusions**

This article contributes to the methodology of measuring a colocation pattern, a fundamental spatial association between objects of different types. Most of existing point-based colocation methods are global measures including the cross K function and the global colocation quotient (CLQ) which assume that such spatial association would not vary in space. Most recently, a local estimate such as the local colocation quotient (LCLQ) is proposed. Our research advances this line of work by developing a simulation-based statistical test for the LCLQ so that each location with an identified spatial association is assessed with a statistical confidence.

Our research indicates that the cross K function tends to over-diagnose the spatial colocation in human activities, regardless of distance metric, because of the lack of control for the joint population distribution. Network distance is generally preferred to Euclidean distance when implementing the global or local CLQ analysis, especially in rural or other areas with a sparse road network or uneven road density. A larger bandwidth leads to a stronger smoothing effect but less local variability. One needs to strike a balance between the two in choosing an appropriate bandwidth. Our case study illustrates the method by examining the colocation between different crimes and various categories of facilities in a city in southern China. Other applications can certainly benefit from the method of colocation analysis of point data.

**Acknowledgments**

The support of Economic Development Assistantship (EDA) from the Graduate School of Louisiana State University to Hu is gratefully acknowledged.